\documentclass[11pt, a4paper]{article}
\usepackage[utf8]{inputenc}
\usepackage[T1]{fontenc}
\usepackage{amsmath, amssymb, amsthm}
\usepackage{graphicx}
\usepackage{geometry}
\usepackage{natbib} 
\usepackage{hyperref} 

\newcommand{\cindep}{\mathrel{\perp\!\!\!\perp}}

\theoremstyle{plain}

\theoremstyle{definition}

\title{Single World Intervention Graphs as Distributions: A Framework for Causal Identification}
\author{Christian Bartels \\ Allschwil, Switzerland \\ 
\texttt{hanschristian.bartels@gmail.com}}
\date{\today} 

\begin{document}

\maketitle

\noindent \textbf{Keywords:} Causal inference, Identifiability, SWIG, causal diagrams \\
\noindent \textbf{MSC Classification:} 62D20

\begin{abstract}
Causal inference seeks to estimate the effect of an intervention on an outcome using observed data, typically via Rubin's potential-outcome framework or Pearl's do-calculus. Following section 9 of Richardson and Robins (2013), this essay treats single-world intervention graphs (SWIGs) as representations of both the observed-data distribution and the interventional distribution, rather than as a bridge to potential outcomes. We demonstrate that this perspective provides a systematic way to derive identifying expressions for estimands defined by interventions on selected variables. Back-door derivations mirror those in existing literature, while front-door derivations offer a distinct pathway that extends more readily to complex settings. Conceptually, the method is simultaneously related to and distinct from Rubin's framework and Pearl's calculus.
\end{abstract}

\section{Introduction}

This essay follows up on section 9 of Richardson and Robins 
\cite{richardson2013single},
which suggests considering single world intervention graphs (SWIGs) as representing the distribution of the observed data and the distribution upon intervention, rather than a relation between intervention and potential outcomes. This view of SWIGs is closely related to do-calculus
\citep{pearl2018book, pearl2000models}.
The essay starts by describing the problem, providing notation, and discussing some basic properties. Building on this, we go through four examples, define estimands, show the assumptions in a SWIG, and demonstrate how the estimands can be identified in the observed data. The first two examples use the SWIG in Figure \ref{swig1} and give the well-known backdoor and front door adjustment formula. The derivations in the appendix consider backdoor and front-door adjustment for longitudinal systems consisting of a series of interventions and observations.

\begin{figure}[h]
\centering
\includegraphics[width=0.6\textwidth]{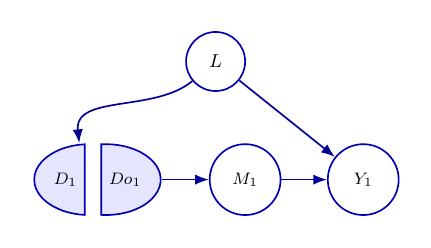}
\caption{Simple SWIG. $D_1$ is the target of intervention, ${Do}_1$ the intervention, $Y_1$ the outcome of interest, L a baseline covariate, and $M_1$ a mediator.}
\label{swig1}
\end{figure}

\section{Methods}

\subsection{Notation}

Having observed data on subjects or patients, we are interested in evaluating the outcome with a hypothetical intervention that forces everyone to take the treatment according to a particular regimen of interest. Both the distribution of the observed data and the distribution of the data upon intervention are described by probability distributions on a set of variables. If there is only a single intervention such as in Figure \ref{swig1}, $q_0(.)$ and $q_1(.)$ denote the distributions of the observed data and the data upon intervention, respectively. With multiple interventions such as in Figure \ref{swig2}, there exist combinatorically many distributions, which can be denoted by $q_s(.)$ with the subscript distinguishing between different combinations of interventions. The derivations explored here select an ordering of the interventions that respects the partial ordering imposed by all conditional independences assumed to be valid. The independences could be represented as causal graph such as in Figures \ref{swig1} or \ref{swig2}. With an ordering of the intervention,  $q_t(.)$ refers to the distribution with interventions up to index t. We refer to the index $t$ also as time. $q_0(.)$ refers to the observed data with no interventions. With a total of n intervention, $q_n(.)$ refers to the distribution with all interventions which is related to the target of estimation – the estimand of interest.

All distributions are defined on the same set of variables
irrespective of interventions. E.g., $Y_n$ denotes
the outcome of interests after the last intervention. Particular 
variables are the targets of the intervention and the 
interventions, $D_t$ and ${Do}_t$, respectively, with the time subscript differentiating subsequent interventions. By
definition, the different distributions $q_t(.)$ with different subscripts differ with respect to the 
interventions. In the observed data and whenever there is no 
intervention at time $t$, ${Do}_t$ is set to the same value as
 $D_t$, i.e., $q_0(D_t,{Do}_t )=\delta(D_t,{Do}_t)$ - as proposed in 
 section 9 of \cite{richardson2013single}. Upon intervention ${Do}_t$ is set to represent the intervention of interest; this essay considers only 
static interventions with ${Do}_t=d_t$ set to fixed values. Dynamic 
strategies that set it to a stochastic function of prior variables 
could be considered. As a consequence of the intervention, 
$D_t$ and ${Do}_t$ are independent $q_n(D_t,{Do}_t)=q_n(D_t ) q_n ({Do}_t)$ or $D_t \cindep {Do}_t$ in $q_n(.)$, in all distributions 
$n \geqslant t$ with intervention at $t$. In general and as for the variables 
representing the intervention, independences and conditional 
independences differ between the distributions representing 
different sets of interventions.

Over-bars together with subscripts denote time series from the first time point to the time of the subscript, e.g., $\overline{D}_t$, is the time series of the target of interventions from time one to time $t$ and $\overline{D}_n$ refers to all interventions, and $\overline{D}_0=\overline{D}_{(i-1)}$ with $i=1$ is an empty set that could be dropped but may be maintained for ease of notation.

Subscripts denote the ordering of all variables and not only the interventions. We referred already to $Y_n$, the outcome at the end of the study after all interventions. Similarly, $Y_t$ denotes the outcome at or after time t and before time $t+1$. For baseline covariates and other variables that do not depend on any intervention (such as randomization), the subscript may be dropped, e.g., $L_0 \equiv L$ could denote baseline covariates, and $R_0 \equiv R$ randomization. 

In general, and as usually done in do-calculus, interventions are 
conditioned upon. In particular, $q_n (Y_n |\overline{Do}_n=\overline{d}_n )$ denotes the estimand, i.e., the distribution 
of the outcome of interest in the overall population when setting 
the interventions to $\overline{d}_n$. If the dependent variables do 
contain only variables up to time $t$, e.g., $\overline{Y}_t$, and 
the conditioning set does not contain variables after time $t$ then 
interventions after time $t$ can be dropped, e.g., $q_n (Y_t |\overline{Do}_n=\overline{d}_n )=q_n (Y_t | \overline{Do}_t=\overline{d}_t )$. Intuitively by denoting the index t as time, the outcome, $Y_t$ at time $t$ does not depend on interventions ${Do}_j$ at later times $j > t$. Formally, for all $j > t$, $Y_t \cindep {Do}_j$ since $Y_t$ and ${Do}_j$ are d-separated. By construction there are no paths with arrows into ${Do}_j$, and there exists no collider that could introduce correlations, since no variable after ${Do}_j$ is included.
In addition, the distributions representing 
interventions after time $t$ are all the same since they do not 
contain any variable after time $t$, e.g., $q_n (Y_t |\overline{Do}_t=\overline{d}_t )=q_j (Y_t |\overline{Do}_t=\overline{d}_t $) for 
all $j$ with $t \leq j \leq n$. Baseline covariates are the extreme case as 
they do not depend on any intervention, e.g., $q_n (L_0 |\overline{Do}_n=\overline{d}_n )=q_n (L)=q_0 (L)$. If the 
conditioning set contains variables after time $t$, then 
interventions after time $t$ cannot simply be dropped, since the 
conditioning set may contain colliders. This needs to be checked on 
a case-by-case basis.

\subsection{Relation to different existing notations}

The notation introduced in the previous section is similar to other existing notations. Probability distributions like the estimand of interest, $q_n (Y_n |\overline{Do}_n=\overline{d}_n )$, and observed data that could be related to it, $q_0 (Y_n |\overline{D}_n=\overline{d}_n )$, are expressed similarly using potential outcomes \citep{richardson2013single, hernan2020causal} or do-calculus \citep{pearl2018book, pearl2000models}. For the estimand, $q_n (Y_n |\overline{Do}_n=\overline{d}_n )$, the corresponding expressions are $Pr(Y_n^{\overline{D}_n = \overline{d}_n})$ or $Pr(Y_n | do(\overline{d}_n))$, respectively. For the observed data, expressions are almost identical, $q_0 (Y_n |\overline{D}_n=\overline{d}_n )$ or $Pr(Y_n | \overline{D}_n = \overline{d}_n)$. 

\subsection{Consistency assumptions}

Identification of the estimand in the observed data has to rely on consistency assumptions. These assumptions were already implied when introducing the target of intervention and the intervention above. If the two are the same, e.g., if $D_t={Do}_t=d_t$, then distributions that have or do not have an intervention at time $t$ are the same, i.e., 
\begin{alignat*}{1}
&q_t\left(\ldots|D_t={Do}_t=d_t,\ \ldots\right) =q_{t-1}\left(\ldots|D_t={Do}_t=d_t,\ \ldots\right)
\end{alignat*}
If all interventions are the same, then the corresponding distribution can be identified in the observed data, i.e.,
\begin{alignat*}{1}
    &q_t\left(\ldots|{\overline{D}}_t={\overline{Do}}_t={\overline{d}}_t,\ \ldots\right) =q_0\left(\ldots|{\overline{D}}_t={\overline{Do}}_t={\overline{d}}_t,\ \ldots\right) \label{eqcon} \nonumber
\end{alignat*}

\subsection{Conditional independences supporting identification}

Starting from an expression that depends on interventions, there exist two ways to obtain expressions, in which interventions and targets of intervention are the same (which is the prerequisite to use the consistency assumption). The first requires (conditional) independence of the dependent variable and the target of intervention, the second requires (conditional) independence of the dependent variable and the intervention. Illustrating this for the estimand, $q_n\left(Y_n|{\overline{Do}}_n={\overline{d}}_n\right)$, and assuming that we need to condition on variables $L$ or $M$, respectively (see also Figure~\ref{swig1}), the two conditional independences are
\begin{equation}
    Y_n\cindep{\overline{D}}_n|L,{\overline{Do}}_n \label{eqbd}
\end{equation}
\begin{equation}
    Y_n\cindep{\overline{Do}}_n|M,{\overline{D}}_n \label{eqfd}
\end{equation}

The first corresponds to exchangeability as discussed in text books (Hernan and Robins JM, 2020). It supports identification via 
\begin{eqnarray}
q_n\left(Y_n|{\overline{Do}}_n={\overline{d}}_n,L\right) 
   & = & q_n\left(Y_n|{\overline{Do}}_n={\overline{d}}_n,{\overline{D}}_n={\overline{d}}_n,L\right) \nonumber\\
   & = & q_0\left(Y_n|{\overline{Do}}_n={\overline{d}}_n,{\overline{D}}_n={\overline{d}}_n,L\right) \nonumber\\
   & = & q_0\left(Y_n|{\overline{D}}_n={\overline{d}}_n,L\right) \nonumber
\end{eqnarray}
Here we used first the conditional independence (Eq. \ref{eqbd}), then consistency, and finally the fact that in the observed data the intervention and the target of intervention are the same. 

The second conditional independence can be used by first introducing the target of intervention using the law of total probability:
\begin{alignat*}{1}
&q_n\left(Y_n \mid \overline{Do}_n = \overline{d}_n, M \right) \\
& \qquad = \sum_{\overline{d'}_n} q_n\left(Y_n \middle| \overline{Do}_n = \overline{d}_n, \overline{D}_n = \overline{d'}_n, M \right) q_n\left(\overline{D}_n = \overline{d'}_n \middle| \overline{Do}_n = \overline{d}_n, M \right) \\
& \qquad = \sum_{\overline{d'}_n} q_n\left(Y_n \middle| \overline{Do}_n = \overline{d'}_n, \overline{D}_n = \overline{d'}_n, M \right) q_n\left(\overline{D}_n = \overline{d'}_n \middle| \overline{Do}_n = \overline{d}_n, M \right) \\
& \qquad = \sum_{\overline{d'}_n} q_0\left(Y_n \middle| \overline{Do}_n = \overline{d'}_n, \overline{D}_n = \overline{d'}_n, M \right) q_n\left(\overline{D}_n = \overline{d'}_n \middle| \overline{Do}_n = \overline{d}_n, M \right) \\
& \qquad = \sum_{\overline{d'}_n} q_0\left(Y_n \middle| \overline{D}_n = \overline{d'}_n, M \right) 
q_n\left(\overline{D}_n = \overline{d'}_n \middle| \overline{Do}_n = \overline{d}_n, M \right)
\end{alignat*}
After having introduced the target of intervention, this proceeds as in the first case, by using the conditional independence, consistency, and finally removing redundant variables in the observed distribution. We may refer to the two conditional independences as backdoor (Eq. \ref{eqbd}) and front-door (Eq. \ref{eqfd}) conditional independence. The backdoor is used routinely in causal inference literature. A minor difference is that here the conditioning on the intervention is explicit whereas when using potential outcome notation it is implicit. 
The front-door conditional independence is new. Do-calculus cannot formulate it as it removes the target of intervention when applying the do operator. The potential outcome notation is not well suited since it does not condition explicitly on the intervention.


\section{Examples}

The next two subsections derive the common back-door and front-door adjustment formula using the definitions from the previous section. The Appendix goes through more involved examples applying back-door and front-door adjustment in a longitudinal setting with multiple interventions. 

\subsection{Back-door adjustment}

Except for minor differences in notation, back-door adjustment proceeds in the same way as within the potential outcome framework. We use the SWIG in Figure \ref{swig1} and define the estimand as $q_1\left(Y_1|{Do}_1=d_1\right)$. The target of estimation, $Y_1$ and the target of intervention $D_1$ are d-separated by conditioning on $L$; the intervention, ${Do}_1$,  can be added to the conditioning set, giving the conditional independence $Y_1 \cindep\ D_1|L,{Do}_1$. We have
\begin{alignat*}{1}
q_1\left(Y_1|{Do}_1=d_1\right) 
&=\sum_{l}{q_1\left(Y_1|{Do}_1=d_1,L=l\right)}q_1\left(L=l|{Do}_1=d_1\right) 		\\
&=\sum_{l}{q_1\left(Y_1|{Do}_1=d_1,D_1=d_1,L=l\right)} q_1\left(L=l|{Do}_1=d_1\right) 	\\
&=\sum_{l}{q_1\left(Y_1|{Do}_1=d_1,D_1=d_1,L=l\right)q_1\left(L=l\right)} \\
&=\sum_{l}{q_0\left(Y_1|{Do}_1=d_1,D_1=d_1,L=l\right)q_0\left(L=l\right)}  \\	
&=\sum_{l}{q_0\left(Y_1|D_1=d_1,L=l\right)}q_0\left(L=l\right)		
\end{alignat*}
The derivation used the law of total probability, conditional independence, removed intervention after the last 
dependent variable, consistency, and finally redundancy of $D_1$ and ${Do}_1$ in the observed data.

\subsection{Front-door adjustment}

Derivation of front-door adjustment can proceed differently than with other frameworks. There is no need to define an intervention on the mediator, nor to consider conditioning on an unobserved confounder. To illustrate this, we use again the SWIG in Figure 1, assuming that the covariate $L$ is unobserved, and focus on the same estimand, $q_1\left(Y_1|{Do}_1=d_1\right)$ as above. Conditioning on $M_1$, the outcome $Y_1$ is d-separated from the intervention ${Do}_1$; adding $D_1$ to the conditioning set does not change this. The latter is the case here but is not guaranteed, since $D_1$ could be a collider. This gives the front door conditional independence, $Y_1\cindep{Do}_1|M,D_1$.

The derivation proceeds by first introducing the target of intervention $D_1$ and the mediator using the law of total probability
\begin{alignat*}{1}
&q_1\left(Y_1|{Do}_1=d_1\right) \\
&=\sum_{m}{q_1\left(Y_1|{Do}_1=d_1,M=m\right)} q_1\left(M=m|{Do}_1=d_1\right) 		\\
&=\sum_{m,{d^\prime}_1} q_1\left(Y_1|{Do}_1=d_1,D_1={d^\prime}_1,M=m\right) q_1\left(D_1={d^\prime}_1|{Do}_1=d_1,M=m\right) q_1\left(M=m|{Do}_1=d_1\right) 
\end{alignat*}
This expression has three terms. The first can be identified using the front door conditional independence. The third and second describe the dependence of the mediator on the intervention and the target of intervention on the intervention. These can be estimated using back-door independences. Based on the SWIG $M_1\cindep\ D_1|{Do}_1$ and $D_1 \cindep\ M_1,{Do}_1$; for both d-separation relies on excluding the collider $Y_1$ from the conditioning set. This gives
\begin{alignat*}{1}
&q_1\left(Y_1|{Do}_1=d_1\right) \\
& \quad =\sum_{m,{d^\prime}_1} q_1\left(Y_1|{Do}_1={d\prime}_1,D_1={d^\prime}_1,M=m\right)  q_1\left(D_1={d^\prime}_1\right)q_1\left(M=m|{Do}_1=d_1,D_1=d_1\right) 
\end{alignat*}
The first and third term can be identified by consistency, the second term since $D_1$ does not depend on any intervention. 
\begin{alignat*}{1}
&q_1\left(Y_1|{Do}_1=d_1\right) \\
& \quad =\sum_{m,{d^\prime}_1} q_0\left(Y_1|{Do}_1={d\prime}_1,D_1={d^\prime}_1,M=m\right) q_0\left(D_1={d^\prime}_1\right)q_0\left(M=m|{Do}_1=d_1,D_1=d_1\right) 
\end{alignat*}
This can be further simplified, since ${Do}_1$ and $D_1$ are redundant in the observed data, giving the well-known front door expression to identify the estimand in the observed data:
\begin{alignat*}{1}
&q_1\left(Y_1|{Do}_1=d_1\right) \\
& \quad =\sum_{m,{d^\prime}_1} q_0\left(Y_1|D_1={d^\prime}_1,M=m\right) q_0\left(D_1={d^\prime}_1\right)q_0\left(M=m|D_1=d_1\right) 
\end{alignat*}
The front door estimator consists of a term describing the effect of the intervention on the mediator and a term of the effect of the mediator on the outcome. The effect of the mediator on the outcome conditions on the observed doses (target of intervention), and the estimator averages over all observed values of the doses.

\section{Discussion and conclusion}

This essay was inspired by section 9 of 
\cite{richardson2013single},
which suggests viewing SWIGs as representations either of the observed data distribution or of the distribution under intervention, rather than as a bridge between interventions and potential outcomes. With this perspective, one can consider a sequence of distributions over the same variables, differing in the nature of the interventions - such as $q_0\left(.\right)$  for observed data and $q_n\left(.\right)$ for interventional data. A natural question arises: how might information from $q_0\left(.\right)$  be used to estimate $q_n\left(.\right)$? 

The resulting viewpoint is related to established causal inference frameworks and, in particular, Rubin’s potential outcome approach 
\citep{holland1986statistics}
and Pearl’s do-calculus 
\citep{pearl2000models}; for comparisons see
\citep{imbens2020potential, pearl2009causal}. 
The approach outlined here shares with do-calculus the distinction between observed and interventional data distributions, though it does not employ the do-operator or its formal machinery. The connection to the potential outcome framework lies in using SWIGs and relying on conditional independences, but without direct reference to potential outcomes.

Beyond what is already existing, the proposed view suggests an alternative way to derive front-door adjustment formulas. Sticking with the example used here, where the primary interest is the effect of intervening on the doses, $D_t$, front door adjustment requires adjusting on these doses. There exist now at least three ways to do derive this.

First, the traditional method (see also Appendix) introduces an intervention on the mediator. Although this mediator intervention is not of primary interest, it is considered because the effect of the mediator on the outcome can be adjusted for the doses (using back-door adjustment).

Second, \citep{hernan2020causal}, Technical Point 21.11 outlines an approach where one conditions on an unobserved confounder to identify the estimand in observed data. This method is useful since front-door adjustment is needed in situations with unobserved confounding. If one could observe all relevant variables, standard back-door adjustment would suffice. Hernan refers to this in Section 21.6 as the big g-formula. Once the estimand is identified in observed data, the unobserved variable can be removed through a process of refactorization that relies on the observed doses. This kind of refactoring has been demonstrated for simpler systems (such as Figure \ref{swig1}), though its feasibility in more complex scenarios (like Figure \ref{swig2}, with sequential conditional independences) is less certain.

Here, we introduced a third possibility, using the particular view of SWIG together with front-door conditional independence such as Eq. (\ref{eqfd}), $Y_n\cindep{\overline{Do}}_n|M,{\overline{D}}_n$, that expresses an independence between the outcome and the intervention when conditioning on the mediator. The target of intervention ${\overline{D}}_n$ are part of the SWIG when viewed as representing the distribution relevant to the estimand. The target of intervention ${\overline{D}}_n$ can be introduced using the law of total probability irrespective of any conditional independences. To identify the estimand in the observed data, we need expressions that condition on the target of intervention and the intervention, and the two variables are set to the same value, i.e., ${\overline{D}}_t={\overline{Do}}_t={\overline{d}}_t$. Setting them to the same value is possible using either front-door or backdoor conditional independences (Eqs. \ref{eqbd} and \ref{eqfd}), which are symmetric in this sense.

The newly introduced approach does not call for interventions on mediators, but instead uses only the interventions specified by the estimand. This eliminates the need for hypothetical auxiliary interventions on mediators, streamlining the identification process in longitudinal settings with unobserved confounding and may offer considerable advantages for more complex systems and arbitrary estimands.

Two general strategies for assessing identifiability can be considered: a top-down and bottom-up approach. In the top-down strategy, one begins with the estimand and introduces the variables required for its identification. Alternatively, the bottom-up approach works by recursively checking which variables can be identified by conditioning only on variables already established.

In summary, using SWIGs as direct representations of observed or interventional data offers an interesting, practical and flexible viewpoint, especially for developing algorithms that assess causal identifiability.

\begin{appendix}
\section{Additional examples}\label{appn} 

\subsection{Longitudinal with sequential conditional independences}

\begin{figure*}[h]
\centering
\includegraphics[width=0.8\textwidth]{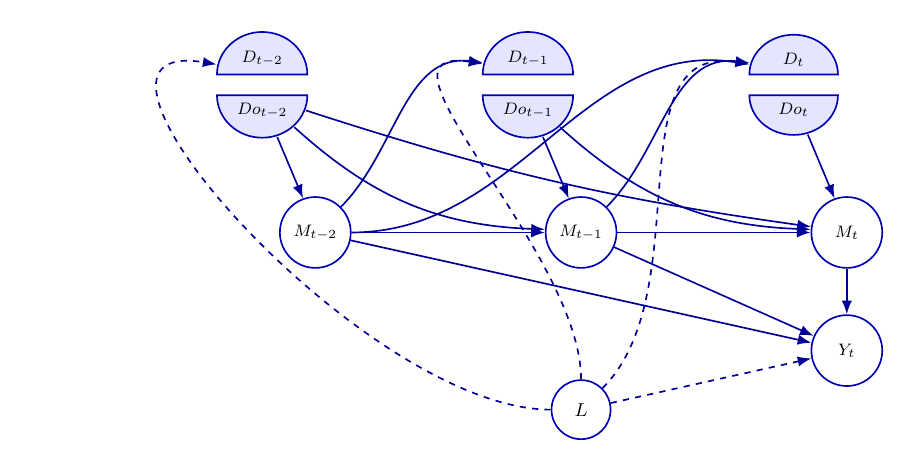}
\caption{Longitudinal SWIG. $D_t$ are targets of intervention, ${Do}_t$ interventions, $M_t$ mediators, $Y_t$ the outcome of interest at the end of the study, $L$ a baseline covariate.}
\label{swig2}
\end{figure*}

Figure \ref{swig2} shows the SWIG for longitudinal example used in this and subsequent sections. This section, focuses on estimating the effect of the intervention on the mediator. In the subsequent section it is the effect on the final outcome. The estimand in this section is $q_n({\overline{M}}_n|{\overline{Do}}_n={\overline{d}}_n\ )$.

The identification of the estimand in the observed data relies on backdoor-like sequential conditional independence $M_t\cindep{\overline{D}}_t|{\overline{M}}_{t-1},{\overline{Do}}_t$. To apply it,
\begin{equation}\label{eq:seq_cond}
\begin{alignedat}{1}
q_n({\overline{M}}_n|{\overline{Do}}_n={\overline{d}}_n\ &=\prod_{t}{q_n\left(M_t|{\overline{M}}_{t-1},{\overline{Do}}_n={\overline{d}}_n\right)} 	\\
&=\prod_{t}{q_n\left(M_t|{\overline{M}}_{t-1},{\overline{Do}}_t={\overline{d}}_t\right)} 	\\
&=\prod_{t}{q_n\left(M_t|{\overline{M}}_{t-1},{\overline{Do}}_t={\overline{d}}_t,{\overline{D}}_t={\overline{d}}_t\right)} 		\\	
&=\prod_{t}{q_0\left(M_t|{\overline{M}}_{t-1},{\overline{Do}}_t={\overline{d}}_t,{\overline{D}}_t={\overline{d}}_t\right)} 		\\	
&=\prod_{t}{q_0\left(M_t|{\overline{M}}_{t-1},{\overline{D}}_t={\overline{d}}_t\right)}. 			
\end{alignedat}
\end{equation}
Here, we used the product rule, removed interventions after the last variable, used conditional independence, consistency, and removed redundant intervention variables in the observed data.

One option to estimate the resulting expression is referred to as g-computation: Models are built for subsequent time points. Each model conditions on earlier outcomes. After having estimated them based on the observed data, they can be used sequentially to sample outcomes starting from the first time point and extending to the last time point.

\subsection{Longitudinal with front door}

To illustrate front door adjustment in the longitudinal setting, we focus on $q_n(Y|{\overline{Do}}_n={\overline{d}}_n\ )$ as estimand and assume that the covariate L is unobserved. To facilitate notation, all relevant variables are introduced as a first step, and subsequent steps focus on identifying the resulting joint distribution. With the law of total probability, we get
\begin{alignat*}{1}
q_n\left(Y\middle|{\overline{Do}}_n={\overline{d}}_n\ \right) &=\sum_{{\overline{d\prime}}_n,{\overline{m}}_n\ }{q_n(Y,{\overline{D}}_n={\overline{d\prime}}_n,{\overline{M}}_n={\overline{m}}_n\ |{\overline{Do}}_n={\overline{d}}_n)\ }
\end{alignat*}
Using the product rule, the joint distribution is split into a series of terms, we do it in two steps for ease of reading, first,
\begin{alignat*}{1}
&q_n(Y,{\overline{D}}_n={\overline{d\prime}}_n,{\overline{M}}_n={\overline{m}}_n\ |{\overline{Do}}_n={\overline{d}}_n)=	\\
&\qquad q_n(Y\ |{\overline{Do}}_n={\overline{d}}_n,{\overline{D}}_n={\overline{d\prime}}_n,{\overline{M}}_n={\overline{m}}_n) \\
&\qquad \times \prod_{t} q_n(D_t={d\prime}_t,M_t=m_t\ | {\overline{Do}}_n={\overline{d}}_n,{\overline{D}}_{t-1}={\overline{d\prime}}_{t-1},{\overline{M}}_{t-1}={\overline{m}}_{t-1})
\end{alignat*}
then,
\begin{alignat*}{1}
&q_n(Y,{\overline{D}}_n={\overline{d\prime}}_n,{\overline{M}}_n={\overline{m}}_n\ |{\overline{Do}}_n={\overline{d}}_n)= \\
& \qquad q_n(Y\ |{\overline{Do}}_n={\overline{d}}_n,{\overline{D}}_n={\overline{d\prime}}_n,{\overline{M}}_n={\overline{m}}_n) \\
& \qquad \times \prod_{t}{q_n(M_t=m_t\ |} {\overline{Do}}_n={\overline{d}}_n,{\overline{D}}_t={\overline{d\prime}}_t,{\overline{M}}_{t-1}={\overline{m}}_{t-1}) \\
& \qquad \times q_n(D_t={d\prime}_t\ | {\overline{Do}}_n={\overline{d}}_n,{\overline{D}}_{t-1}={\overline{d\prime}}_{t-1},{\overline{M}}_{t-1}={\overline{m}}_{t-1})
\end{alignat*}

This has now a series of terms that describe the response, the mediator and the target of intervention. Each of these terms needs to be identified in the observed data. For the response, we can use front-door type conditional independence, $Y\cindep{\overline{Do}}_n\ |{\overline{D}}_n,{\overline{M}}_n$
\begin{alignat*}{1}
q_n\left(Y\ \middle|{\overline{Do}}_n={\overline{d}}_n,{\overline{D}}_n={\overline{d^\prime}}_n,{\overline{M}}_n={\overline{m}}_n\right) 
&=q_n\left(Y\ \middle|{\overline{Do}}_n={\overline{d\prime}}_n,{\overline{D}}_n={\overline{d^\prime}}_n,{\overline{M}}_n={\overline{m}}_n\right) \\
& =q_0\left(Y\ \middle|{\overline{Do}}_n={\overline{d}\prime}_n,{\overline{D}}_n={\overline{d^\prime}}_n,{\overline{M}}_n={\overline{m}}_n\right) \\
& =q_0\left(Y\ \middle|{\overline{D}}_n={\overline{d^\prime}}_n,{\overline{M}}_n={\overline{m}}_n\right)
\end{alignat*}
This is similar to the case for n=1 discussed above, except that the conditioning happens on the entire time series of the mediator and the target of intervention

For the mediator, the derivation is the same as in the previous section. We got
\begin{alignat*}{1}
&q_n\left(M_t=m_t\ \middle|{\overline{Do}}_n={\overline{d}}_n,{\overline{D}}_t={\overline{d^\prime}}_t,{\overline{M}}_{t-1}={\overline{m}}_{t-1}\right) = q_0\left(M_t|{\overline{M}}_{t-1},{\overline{D}}_t={\overline{d}}_t\right) 
\end{alignat*}

For the doses, we require sequential front door independence $D_t\cindep{\overline{Do}}_{t-1}|{\overline{M}}_{t-1},{\overline{D}}_{t-1}$. This independence holds since all paths connecting ${\overline{Do}}_{t-1}$ and $D_t$ pass through ${\overline{M}}_{t-1}$. 
\begin{alignat*}{1}
&q_n\bigl(D_t={d\prime}_t\ \big|{\overline{Do}}_n={\overline{d}}_n,{\overline{D}}_{t-1}={\overline{d^\prime}}_{t-1},{\overline{M}}_{t-1}={\overline{m}}_{t-1}\bigr) \\
& \qquad =q_n\bigl(D_t={d\prime}_t\ \big|  {\overline{Do}}_{t-1}={\overline{d}}_{t-1},{\overline{D}}_{t-1}={\overline{d^\prime}}_{t-1},{\overline{M}}_{t-1}={\overline{m}}_{t-1}\bigr)	\\
& \qquad =q_n\bigl(D_t={d\prime}_t\ \big|  {\overline{Do}}_{t-1}={\overline{d\prime}}_{t-1},{\overline{D}}_{t-1}={\overline{d^\prime}}_{t-1},{\overline{M}}_{t-1}={\overline{m}}_{t-1}\bigr) \\
& \qquad =q_0\bigl(D_t={d\prime}_t\ \big| {\overline{Do}}_{t-1}={\overline{d\prime}}_{t-1},{\overline{D}}_{t-1}={\overline{d^\prime}}_{t-1},{\overline{M}}_{t-1}={\overline{m}}_{t-1}\bigr) \\
& \qquad =q_0\bigl(D_t={d\prime}_t\ \big| {\overline{D}}_{t-1}={\overline{d^\prime}}_{t-1},{\overline{M}}_{t-1}={\overline{m}}_{t-1}\bigr) 		
\end{alignat*}
Here, we first used that $D_t$ has no dependence on interventions after $t-1$, then used the front-door type conditional independence, then consistency, and finally removed redundant variables from the observed distribution.
This term estimates the distribution of the target of intervention at time $t$ that would be observed when intervening on prior time points setting ${\overline{Do}}_{t-1}={\overline{d}}_{t-1}$. The intervention does not occur in the expression but is mediated via ${\overline{M}}_{t-1}={\overline{m}}_{t-1}$.

Putting the three identified terms together gives
\begin{equation}\label{eq:seq_front}
\begin{alignedat}{1}
&q_n(Y,{\overline{D}}_n={\overline{d\prime}}_n,{\overline{M}}_n={\overline{m}}_n\ |{\overline{Do}}_n={\overline{d}}_n)= \\
& \quad q_0\left(Y\ \middle|{\overline{D}}_n={\overline{d^\prime}}_n,{\overline{M}}_n={\overline{m}}_n\right)\times \\
& \quad	\prod_{t}{q_0\left(M_t=m_t|{\overline{M}}_{t-1}={\overline{m}}_{t-1},{\overline{D}}_t={\overline{d}}_t\right)} q_0\left(D_t={d\prime}_t\ \middle|{\overline{D}}_{t-1}={\overline{d^\prime}}_{t-1},{\overline{M}}_{t-1}={\overline{m}}_{t-1}\right) .
\end{alignedat}
\end{equation}
With n=1, this simplifies to the well-known front-door adjustment formula discussed above.

\subsection{Intervention on mediator}

\begin{figure*}[h]
\centering
\includegraphics[width=0.8\textwidth]{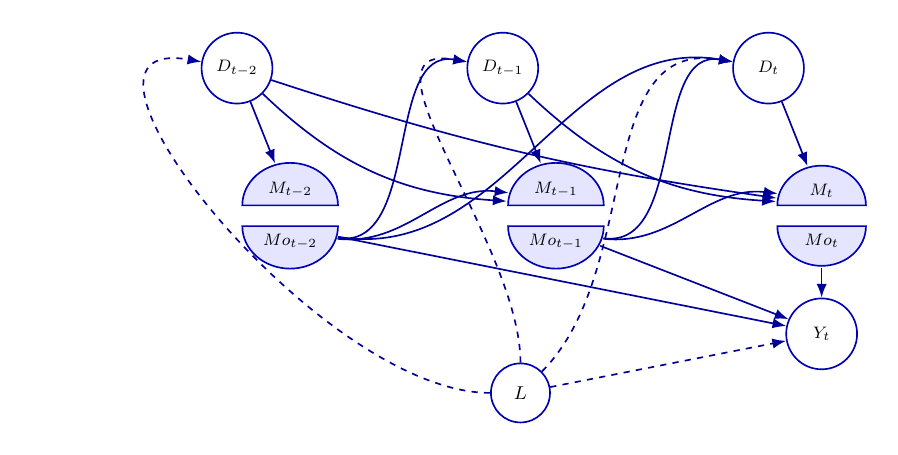}
\caption{Longitudinal SWIG with intervention on mediator. The doses $D_t$ are outcomes. $M_t$ are the targets of intervention, ${Mo}_t$ interventions, $Y_t$ the outcome of interest at the end of the study, $L$ a baseline covariate.}
\label{swig3}
\end{figure*}

For front door adjustment, an alternative to identify the estimand in the observed data is to consider separately the effect of intervening on the doses, and the effect of intervening on the mediator. This is kind of the standard to derive front door adjustment and can be expressed within the potential outcome framework and with do-calculus. Figure \ref{swig2} shows the SWIG of the intervention on the doses. Above in Eq.~\eqref{eq:seq_cond} we derived the effect on the mediator as
\begin{alignat*}{1}
&q_n({\overline{M}}_n={\overline{m}}_n|{\overline{Do}}_n={\overline{d}}_n\ ) =\prod_{t}{q_0\left(M_t=m_t|{\overline{M}}_{t-1}={\overline{m}}_{t-1},{\overline{D}}_t={\overline{d}}_t\right)}.
\end{alignat*}
Figure \ref{swig3} shows the SWIG describing the intervention on the mediator. The estimand for the effect of the mediator on the outcome is $q_{n\prime}\left(Y_n\middle|{\overline{Mo}}_n={\overline{m}}_n\right)$. The dash on the subscript of the distribution emphasizes that this is the distribution with intervention on the mediator rather than the distribution with intervention on the doses. Using the effect of the doses on the mediator as (stochastic) intervention on the effect of the mediator on the outcome gives the effect of the doses on the outcome
\begin{alignat*}{1}
&q_n\left(Y\middle|{\overline{Do}}_n={\overline{d}}_n\ \right)= \sum_{{\overline{m}}_n\ }{q_{n\prime}\left(Y_n\middle|{\overline{Mo}}_n={\overline{m}}_n\right)\times q_n({\overline{M}}_n={\overline{m}}_n|{\overline{Do}}_n={\overline{d}}_n\ )\ }
\end{alignat*}

To identify $q_{n\prime}\left(Y_n\middle|{\overline{Mo}}_n={\overline{m}}_n\right)$, we proceed similarly to other derivations that were presented. First, introduce relevant variables, here the targets of intervention, ${\overline{D}}_n$, with the law of total probability
\begin{alignat*}{1}
&q_{n\prime}\left(Y_n\middle|{\overline{Mo}}_n={\overline{m}}_n\right)= \sum_{{\overline{d\prime}}_n\ }{q_{n\prime}(Y_n,{\overline{D}}_n={\overline{d\prime}}_n\ |{\overline{Mo}}_n={\overline{m}}_n)\ }
\end{alignat*}
Subsequently, to keep notation simple, focus on the conditional probability and split into contributions using product rule
\begin{alignat*}{1}
&q_{n\prime}\left(Y_n,{\overline{D}}_n={\overline{d^\prime}}_n\ \middle|{\overline{Mo}}_n={\overline{m}}_n\right) \\
& \quad =q_{n\prime}(Y\ |{\overline{Mo}}_n={\overline{m}}_n,{\overline{D}}_n={\overline{d\prime}}_n) q_n({\overline{D}}_n={\overline{d\prime}}_n\ |{\overline{Mo}}_{n\prime}={\overline{m}}_n)  \\
& \quad =q_{n\prime}(Y\ |{\overline{Mo}}_n={\overline{m}}_n,{\overline{D}}_n={\overline{d\prime}}_n) \prod_{t}{q_{n\prime}(D_t={d\prime}_t\ |{\overline{Mo}}_n={\overline{m}}_n,{\overline{D}}_{t-1}={\overline{d\prime}}_{t-1})}
\end{alignat*}

The resulting expression describing the doses, $D_i$, can be simplified by removing interventions after the last time-point of the dependent variable. The situation is special in that $D_t$ does not depend on, ${Mo}_t$. The intervention at time $t$, ${Mo}_t$, can be removed as well:
\begin{alignat*}{1}
& q_{n\prime}\left(Y_n,{\overline{D}}_n={\overline{d^\prime}}_n\ \middle|{\overline{Mo}}_n={\overline{m}}_n\right)= \\
& \quad q_{n\prime}\left(Y_n\ \middle|{\overline{Mo}}_n={\overline{m}}_n,{\overline{D}}_n={\overline{d^\prime}}_n\right) \prod_{t}{q_{n\prime}(D_t={d\prime}_t\ |{\overline{Mo}}_{t-1}={\overline{m}}_{t-1},{\overline{D}}_{t-1}={\overline{d\prime}}_{t-1})}
\end{alignat*}
We can then invoke backdoor conditional independences for $Y_n$ and $D_t$ with $\{ {\overline{D}}_n, {\overline{Mo}}_n \}$ and $\{ {\overline{D}}_{t-1}, {\overline{Mo}}_{t-1} \}$ being  sufficient adjustment sets, respectively. For $Y_n$, we have $Y_n\cindep{\overline{M}}_n|{\overline{Mo}}_n,{\overline{D}}_n$; conditioning on ${\overline{D}}_n$ and ${\overline{Mo}}_n$ blocks all confounding paths into ${\overline{M}}_n$ including those induce by $L$. Similarly, for $D_t$, we have $D_t\cindep{\overline{M}}_{t-1}|{\overline{Mo}}_{t-1},{\overline{D}}_{t-1}$; all paths into ${\overline{M}}_{t-1}$ ar blocked by conditioning on ${\overline{D}}_{t-1}$ and ${\overline{Mo}}_{t-1}$. Using the independences, consistency, noting that the observed data does not depend on the intervention that is considered, i.e., $q_0(.)=q_{0\prime}(.)$, and then removing variables that are redundant in the observed data gives
\begin{alignat*}{1}
 q_{n\prime}&\left(Y_n,{\overline{D}}_n={\overline{d^\prime}}_n\ \middle|{\overline{Mo}}_n={\overline{m}}_n\right) \\
& =q_{n\prime}\left(Y_n\ \middle|{\overline{Mo}}_n={\overline{m}}_n,{\overline{D}}_n={\overline{d^\prime}}_n,{\overline{M}}_n={\overline{m}}_n\right) \\
& \quad \times \prod_{t} q_{n\prime}(D_t={d\prime}_t\ |{\overline{Mo}}_{i-1}={\overline{m}}_{i-1}, {\overline{D}}_{t-1}={\overline{d\prime}}_{t-1},{\overline{M}}_{t-1}={\overline{m}}_{t-1})  \\
& =q_0\left(Y_n\ \middle|{\overline{Mo}}_n={\overline{m}}_n,{\overline{D}}_n={\overline{d^\prime}}_n,{\overline{M}}_n={\overline{m}}_n\right) \\
& \quad \times \prod_{t} q_0(D_t={d\prime}_t\ |{\overline{Mo}}_{i-1}={\overline{m}}_{i-1}, {\overline{D}}_{t-1}={\overline{d\prime}}_{t-1},{\overline{M}}_{t-1}={\overline{m}}_{t-1})  \\
& =q_0\left(Y_n\ \middle|{\overline{M}}_n={\overline{m}}_n,{\overline{D}}_n={\overline{d^\prime}}_n\right) \\
& \quad \times \prod_{t} q_0(D_t={d\prime}_t\ |{\overline{M}}_{i-1}={\overline{m}}_{i-1}, {\overline{D}}_{t-1}={\overline{d\prime}}_{t-1}) 
\end{alignat*}
Together with Eq.~\eqref{eq:seq_cond}, this is the same result as in Eq.~\eqref{eq:seq_front}
that has been obtained in the previous section using front-door independences without considering an intervention on the mediator.

\end{appendix}

\section*{Acknowledgments}
I thank my colleagues for valuable discussions on causal inference over the past several years, and Novartis Pharma AG for providing a stimulating working environment. This work was conducted independently of Novartis, and the views expressed are my own.

\bibliographystyle{unsrt}
\bibliography{identify.0.1}

\end{document}